\documentclass[camera]{jpaper}

\usepackage[normalem]{ulem}
\usepackage{fancyhdr}
\usepackage{graphicx}
\usepackage{colortbl}
\usepackage{enumitem}
\usepackage{lipsum}
\usepackage{amsmath}
\usepackage{subfig}
\usepackage{tikz}
\usepackage{color}
\usepackage{soul}
\usepackage{setspace}
\usepackage{cite}
\usepackage{multirow}
\usepackage{tablefootnote}
\usepackage{threeparttable}
\usepackage{xspace}
\usepackage{listings}
\usepackage[noend]{algpseudocode}
\usepackage{mathtools}
\usepackage{MnSymbol,wasysym}
\usepackage[short,nodayofweek,level,12hr]{datetime}

\usepackage{makecell}
\newcolumntype{?}[1]{!{\vrule width #1}}

\usepackage{xurl}
\usepackage{hyperref}

\def\BibTeX{{\rm B\kern-.05em{\sc i\kern-.025em b}\kern-.08em
    T\kern-.1667em\lower.7ex\hbox{E}\kern-.125emX}}

\DeclareRobustCommand\circledblack[1]{\tikz[baseline=(char.base)]{
           \node[shape=circle,draw,inner sep=0pt,fill=black, text=white] (char) {#1};}}

\makeatletter
\def\BState{\State\hskip-\ALG@thistlm}
\makeatother

\newcommand{\mechanism}{{WoLFRaM}\xspace}

\definecolor{dkgreen}{rgb}{0,0.6,0}
\definecolor{gray}{rgb}{0.5,0.5,0.5}
\definecolor{mauve}{rgb}{0.58,0,0.82}
\definecolor{chocolate}{rgb}{0.48, 0.25, 0.0}

\newif\ifsubmission
\submissionfalse

\ifsubmission
\newcommand\lois[1]{#1}

\newcommand\joao[1]{#1}
\newcommand\suyash[1]{#1}
\newcommand\todo[1]{}
\else
\newcommand\lois[1]{{\color{black}#1}} 
\newcommand{\joao}[1]{{\color{black}#1}} 
\newcommand{\joaotwo}[1]{{\color{black}#1}} 
\newcommand{\onur}[1]{{\color{black}#1}} 
\newcommand{\onurtwo}[1]{{\color{black}#1}} 
\newcommand{\onurthree}[1]{{\color{black}#1}} 
\newcommand{\onurfour}[1]{{\color{black}#1}} 
\newcommand{\onurfive}[1]{{\color{black}#1}} 
\newcommand{\onursix}[1]{{\color{black}#1}} 
\newcommand{\onurseven}[1]{{\color{black}#1}} 
\newcommand{\onureight}[1]{{\color{black}#1}} 
\newcommand\suyash[1]{{\color{black}#1}} 
\newcommand\todo[1]{{\color{red} {\bf \fbox{TODO: }}{\it#1}}}
\fi

\begin{document}

\title{\vspace{-0.5cm}\mechanism{}: Enhancing Wear-Leveling and Fault Tolerance  \\ in Resistive Memories \onur{using} Programmable Address Decoders}


\date{}

\newcommand{\affilETH}[0]{\textsuperscript{\S}}
\newcommand{\affilTech}[0]{\textsuperscript{$\ddagger$}}
\newcommand{\affilTexas}[0]{\textsuperscript{$\dagger$}}
\newcommand{\affilRoorkee}[0]{\textsuperscript{$\star$}}

\author{
{Leonid~Yavits\affilTech}\qquad~~~%
{Lois~Orosa\affilETH}\qquad~~~%
{Suyash~Mahar\affilRoorkee}\qquad~~~%
\vspace{2pt}
{João~Dinis~Ferreira\affilETH}\\
{Mattan~Erez\affilTexas}\qquad~~~%
{Ran~Ginosar\affilTech}\qquad~~~%
\vspace{6pt}
\onur{{Onur~Mutlu\affilETH}}\\%
\emph{{\affilTech Technion-Israel Institute of Technology \qquad \affilETH ETH Z{\"u}rich}}\\\emph{{\affilRoorkee Indian Institute of Technology Roorkee \qquad \affilTexas University of Texas Austin%
}}%
}

\definecolor{MidnightBlue}{rgb}{0.1, 0.1, 0.44}
\newcommand{\versionnum}[0]{6.5~---~\today~@~\currenttime~CET} 
\newif\ifcameraready
\camerareadytrue

\maketitle

\fancyhead{}
\ifcameraready
 \thispagestyle{plain}
 \pagestyle{plain}
\else
 \fancyhead[C]{\textcolor{MidnightBlue}{\emph{ICCD 2020 Camera Ready Version \versionnum}}}
 \fancypagestyle{firststyle}
 {
   \fancyhead[C]{\textcolor{MidnightBlue}{\emph{ICCD 2020 Camera Ready Version \versionnum}}}
   \fancyfoot[C]{\thepage}
 }
 \thispagestyle{firststyle}
 \pagestyle{firststyle}
\fi


\begin{abstract}


Resistive memories \onur{have} limited lifetime caused by \lois{limited} write endurance and highly \lois{non-uniform} write access patterns. Two main techniques to mitigate \onur{endurance-related} memory failures are 1) \onurtwo{wear-leveling,} to evenly distribute the writes across the \lois{entire} memory, and 2) \onurtwo{fault tolerance,} to correct memory cell failures.  However, one of the main open challenges in extending the lifetime of existing resistive memories is to make both techniques work together seamlessly and efficiently.

To address this challenge, we propose \mechanism{}, a new mechanism that combines both wear-leveling and fault tolerance techniques at low cost by using a programmable resistive address decoder (PRAD). The key idea of \mechanism{} is to use PRAD for implementing 1) a new efficient wear-leveling mechanism that remaps write accesses to random physical locations on the fly, and 2) \lois{a} new efficient fault tolerance mechanism that recovers from faults by remapping failed memory blocks to available physical locations. \onur{Our evaluations show} that, for a \lois{Phase Change Memory (PCM)} \onurtwo{based system} with cell endurance of $\text{10}^\text{8}$ writes, \mechanism{} increases the memory lifetime by 68\% compared to a \onur{baseline} that implements the best state-of-the-art wear-leveling and fault correction mechanisms. \mechanism{}’s average / worst-case performance and energy overheads are 0.51\% / 3.8\% and 0.47\% / 2.1\% respectively.

\end{abstract}

\section{Introduction}

Resistive memories provide significant advantages over DRAM in terms of non-volatility and \onur{technology} scaling~\cite{kultursay2013evaluating,qureshi2009scalable,wong2010phase,raoux2008phase,yoon2014efficient,Lee2009,meza2013case,lee2010phasescale,zhou2009durable}. However, the limited write endurance of resistive memories, \lois{e.g.,} $10^6$-$10^8$ writes per memory cell in \lois{Phase Change Memory (PCM)}~\cite{han2015enhanced,zhou2016increasing,Lee2009,lee2010phase}, limits their usage as main memory. Workloads with significant non-uniformity in write access patterns can cause early failures in rows that are heavily written, which decreases the expected memory lifetime~\cite{qureshi2009enhancing,qureshi2011practical,Lee2009}. To increase \onur{lifetime}, resistive memories implement wear-leveling \onurfour{techniques
~\cite{zhou2009durable,YongsooJoo2010,Seznec2010,qureshi2009enhancing,seong2010security,qureshi2011practical,Yu2014,JoosungYun2012,Fan2014,han2015enhanced,zhou2016increasing,Zhang2017,Chen2012,Chang2016,Long2013,DuoLiu2014,Pan2016,AghaeiKhouzani2014,Im2014,Cheng2016}}
for leveling out the write non-uniformity by \lois{remapping} frequent writes to less heavily written locations. Unfortunately, memory cells can have very different write endurance \onurtwo{levels} due to process variation, which makes wear-leveling more challenging.

Existing wear-leveling techniques have four drawbacks. First, many early proposals do not consider the write patterns of applications \lois{when remapping} write accesses. As a result, some applications may wear memory out much \onur{more quickly} than others. Second, some techniques do not consider endurance variation across different memory cells~\cite{qureshi2009enhancing, seong2010security, zhou2009durable}, which \onur{can cause} early memory failures due to the failure of cells with lower write endurance. Third, some existing mechanisms~\cite{AghaeiKhouzani2014,DuoLiu2014,zhou2016increasing} \lois{remap} writes \onur{at a coarse granularity} (e.g., at \onur{the granularity} of pages or even larger memory regions), which \onur{reduces wear-leveling} efficiency. Fourth, most techniques are relatively slow because they level out the write non-uniformity in a sequential fashion.
The goal of wear-leveling is to extend the lifetime of memory cells as much as possible.
%
\onurtwo{Once} memory cells reach their endurance \onur{limits}, resistive memory needs fault tolerance to continue operating. Fault tolerance \onurfour{mechanisms~\cite{schechter2010use,seong2010safer,fan2013aegis,qureshi2011pay,yoon2011free,azevedo2013zombie,melhem2012rdis,tavana2018block,cai2015data,cai2015read,luo2018heatwatch,cai2014neighbor,cai2012flash,luo2018improving,cai2017vulnerabilities,cai2013program,cai2013threshold,cai2012error,capps1999method,ipek2010dynamically,awasthi2012efficient, cai2017error, cai2018errors, kline2020, kline2017sustainable, Li2011,li2019selective, Li2012, luo2015warm, luo2016enabling, nair2013archshield, tavana2017remap}} typically enable recovery from several failed bits per data page or data block. 

\onurtwo{Unfortunately, to our knowledge, there is no technique that combines both wear-leveling and fault tolerance techniques in a seamless way to 1) level out the write non-uniformity and 2) tolerate faults when memory \onurthree{cells reach their} endurance limits. A previous work~\cite{Fan2014} shows that naively combining both techniques can result into the malfunction of the \onurthree{system: a commonly-used} wear-leveling technique stops working seamlessly once the first data block fails and is mapped out, since \onur{the data block's} physical position becomes unavailable as a remapping target~\cite{Fan2014}.}

Our goal in this paper is to 1) mitigate the shortcomings of existing wear-leveling mechanisms, and 2) enable \onur{seamless} and efficient integration of wear-leveling and fault tolerance techniques. To this end, we propose \mechanism{} \onurtwo{({\bf W}ear-{\bf L}eveling and {\bf F}ault tolerance for {\bf R}esistive {\bf M}emories)}, the first \onur{integrated} mechanism that combines both wear-leveling and fault tolerance. The \onur{overarching} key idea of \mechanism{} is to use \lois{a} Programmable Resistive Address Decoder (PRAD)~\cite{yavits2017resistive} to decouple memory addresses from physical memory locations\onur{, which serves as a remapping substrate that seamlessly enables both wear-leveling and fault tolerance}.

\lois{PRAD allows programming arbitrary addresses into an address decoder position (\lois{i.e., a decoder row}), which enables dynamic assignment of addresses to physical \lois{memory} rows. During a memory access, PRAD selects the decoder \onur{row} (and consequently the memory row) where the address matches the stored pattern, similar to tag matching in associative caches. In contrast, conventional address decoders used in random access memories are hardwired, and the address of a memory row is permanently linked to the physical \onur{row} of the decoder.}

\mechanism{} \lois{\emph{wear-leveling} mechanism} periodically \lois{reprograms} the address decoder to \lois{remap a} write address to a different physical \lois{memory} location. \mechanism{} \lois{implements a write-access-pattern-aware mechanism that} remaps \onur{frequently-written} addresses at a higher rate \onur{than infrequently-written} \lois{addresses}. \mechanism{} performs \lois{address} remapping transparently, i.e., the data
\onur{is accessed always with the same memory address while its physical location in the memory device might change.} 
\lois{Compared to state-of-the-art \onur{wear-leveling} mechanisms,} \mechanism{} does not require external address remapping techniques, such as explicit remapping tables~\cite{JoosungYun2012,zhou2009durable}, predictable arithmetic mapping ~\cite{qureshi2009enhancing,seong2010security,zhou2016increasing}, or \lois{page table reprogramming}~\cite{AghaeiKhouzani2014,Cheng2016,Pan2016,Im2014,DuoLiu2014,Long2013,Zhang2009,Hu2010,Hu2011}. Compared to the commercially available \lois{Intel Optane DC Persistent Memory Module (Optane DC PMM)}~\cite{1903.05714}, \mechanism{} does not require a separate DRAM with power failure protection mechanisms for \onurtwo{storing} translation tables \onurtwo{needed for wear-leveling}.

\lois{\mechanism{} \emph{fault tolerance} mechanism simply uses PRAD to remap the address of a failed memory row to an empty memory row \onur{without errors}.}

\lois{We evaluate our proposal using Phase Change Memory (PCM). Our results show that, compared to \lois{a} state-of-the-art two-level Security Refresh wear-leveling \lois{mechanism}~\cite{seong2010security} coupled with \lois{an} $\text{ECP}_1$ failure correction \lois{mechanism}~\cite{schechter2010use}, \mechanism{} achieves 1) 68\% longer lifetime, 2) 0.51\% (3.8\%) average (maximum) performance overhead for \onur{SPEC CPU2006} benchmarks, and 3) 0.47\% (2.1\%) average (worst-case) energy overhead for \onur{SPEC CPU2006} benchmarks.}

This paper makes the following \lois{key} contributions:
\begin{itemize}[leftmargin=*]
\item \lois{We propose \mechanism{}, the first mechanism that seamlessly integrates wear-leveling and fault tolerance into resistive memories by using a Programmable Resistive Address Decoder (PRAD). \mechanism{} overcomes the four main drawbacks \onurtwo{of} previous mechanisms by 1) considering the write patterns of the application, 2) considering endurance \onur{variation} across different memory cells, 3) remapping writes \onur{at fine granularity}, and 4) performing \onur{wear-leveling} \onur{21.7x} faster than \onur{the best state-of-the-art mechanism}.}
\item \lois{We evaluate the lifetime, performance and energy of \mechanism{} compared to a combination of a state-of-the-art \onur{wear-leveling} mechanism~\cite{seong2010security} and a state-of-the-art fault tolerance mechanism~\cite{schechter2010use}. Our results show that \mechanism{} \onur{provides} \suyash{a} significantly longer memory lifetime \onur{at} significantly lower performance and energy overheads.}
\end{itemize}

\section{Background}
\label{sec:background}

We \lois{provide the necessary background on} the organization \lois{and operation} of a \suyash{typical} resistive memory, and the basic operation of a conventional address decoder \lois{and a programmable resistive address decoder}.


\subsection{Resistive Memory Organization}
\begin{figure}[ht] 
\centering
    \includegraphics[width=1.0\linewidth]{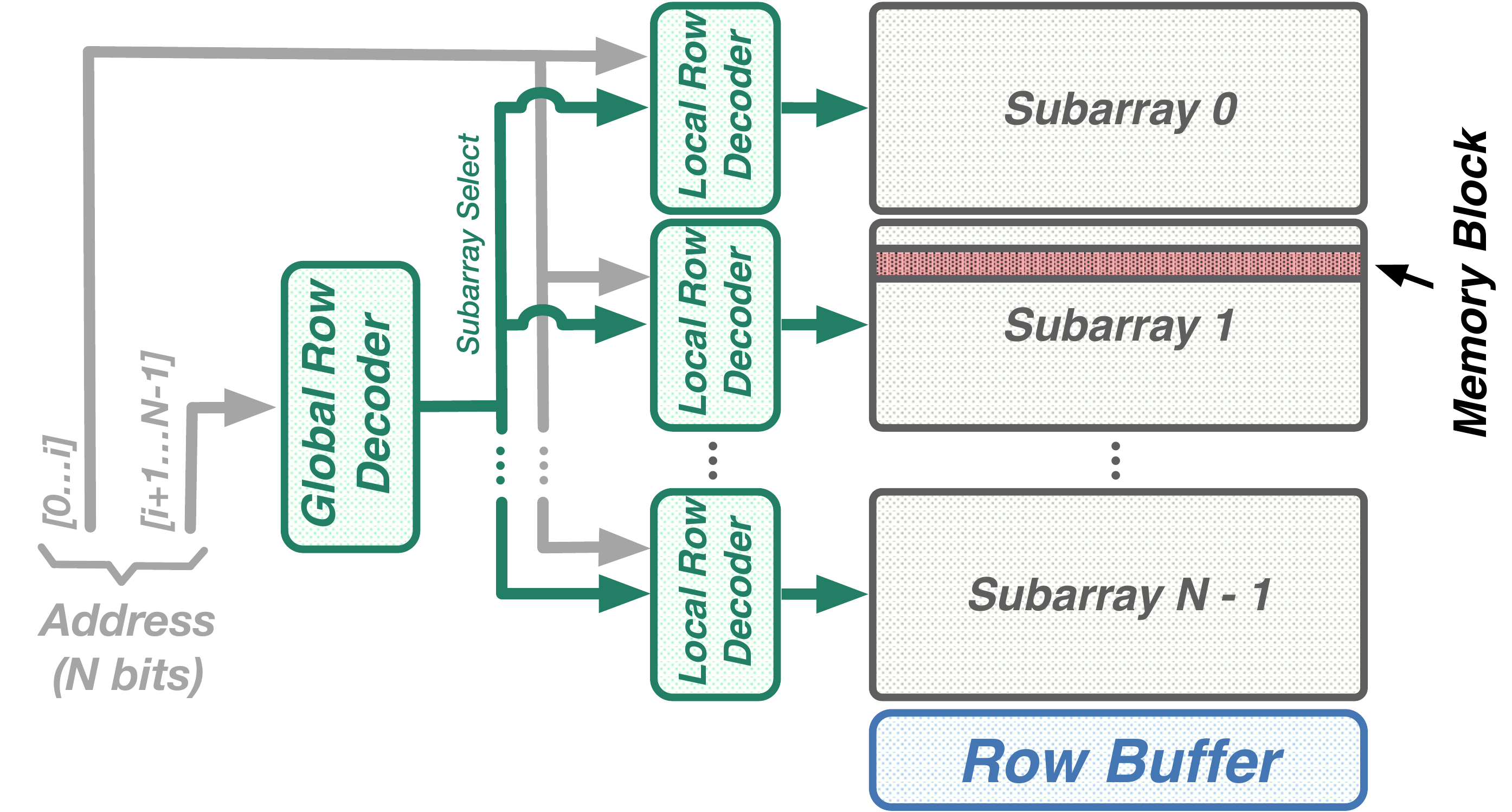}
    \caption{\lois{Overview of a resistive} memory \onur{bank.}}
    \label{fig:memory_bank}
\end{figure}

\lois{A} resistive memory contains multiple independently controlled \onursix{banks~\cite{seong2010safer, Lee2009, Kim2012,meza2012evaluating,seshadri2019dram}}, similar to DRAM. A resistive memory bank (Figure~\ref{fig:memory_bank}) \lois{is composed of} an array of memory \lois{cells} \lois{organized} into multiple subarrays \onurfour{(e.g., \onursix{64-128~\cite{Kim2012,song2019enabling,meza2012evaluating,seshadri2013rowclone,chang2016low,Lee2009,lee2010phasescale,kim2018solar,chang2014improving,seshadri2017ambit,seshadri2015fast,ghose2019demystifying,seshadri2019dram}})} of multiple rows \onurfour{(e.g., \onursix{512-1024~\cite{seong2010safer,Kim2012,meza2012evaluating,seshadri2013rowclone,chang2016low,Lee2009,lee2010phasescale,kim2018solar,chang2014improving,seshadri2017ambit,seshadri2015fast,ghose2019demystifying,seshadri2019dram}})}.

\onur{Bank-level} address decoding is hierarchical. There are typically two \lois{address} decoding levels~\cite{Kim2012}: \lois{1) the} global row decoder selects a subarray, and 2)  \lois{the} local row decoder selects a \onurtwo{row in the subarray} that contains the target memory block. \lois{The target} memory block,
typically 256B to \onurfive{1kB~\cite{seong2010security,Kim2012,meza2012case,meza2012evaluating,Lee2009,chang2016understanding,lee2013tiered,lee2017design}}, is individually addressable within a subarray, while a subarray is individually addressable within a bank. Individual addressing is important since it enables fine-grained remapping of a single data block, as well as a single subarray.
 
\subsection{Resistive Memory Operation}
\label{sec:resistive_memory_operation}
 
To serve a memory request that accesses data at a particular \lois{memory block}, the memory controller issues three commands to a bank. Each command triggers a specific sequence of events within the bank. These \onurfour{commands~\cite{JEDEC,standard2011low,Kim2012,lee2013tiered,lee2015adaptive,lee2015decoupled,Lee2009,meza2012case,meza2012evaluating,song2019enabling,song2020improving}}\onur{,} used commercially and in research~\cite{Lee2009,Kim2012}\onur{,} are similar
to \onur{the} DDRx protocol commands:

\begin{itemize}[leftmargin=3mm,itemsep=0mm,parsep=0mm,topsep=0mm]
    \item {\bf ACT}: an activate command, which reads the memory row \onur{into} the row buffer.
    \item {\bf PRE}: a precharge command, which writes back the contents of the row buffer to a row in the memory \joaotwo{array and} precharges the \onur{bitlines} for the next access~\cite{Lee2009}.\footnote{Unlike \onur{in} DRAM, the row buffer is writen back to the memory array only if the content is modified by a write access~\cite{meza2012evaluating}.} 
    \item {\bf RD/WR}: a read/write command, which reads/writes new data from/to the row buffer. 
\end{itemize}

For more detail and background on the operation of resistive memories, please refer \onurfour{to~\cite{Lee2009,lee2010phasescale,meza2012evaluating,song2019enabling}}.
 
\subsection{Conventional Address Decoder}
\label{sec:conv_dec}

\begin{figure}[h] 
\centering
    \includegraphics[width=1\linewidth]{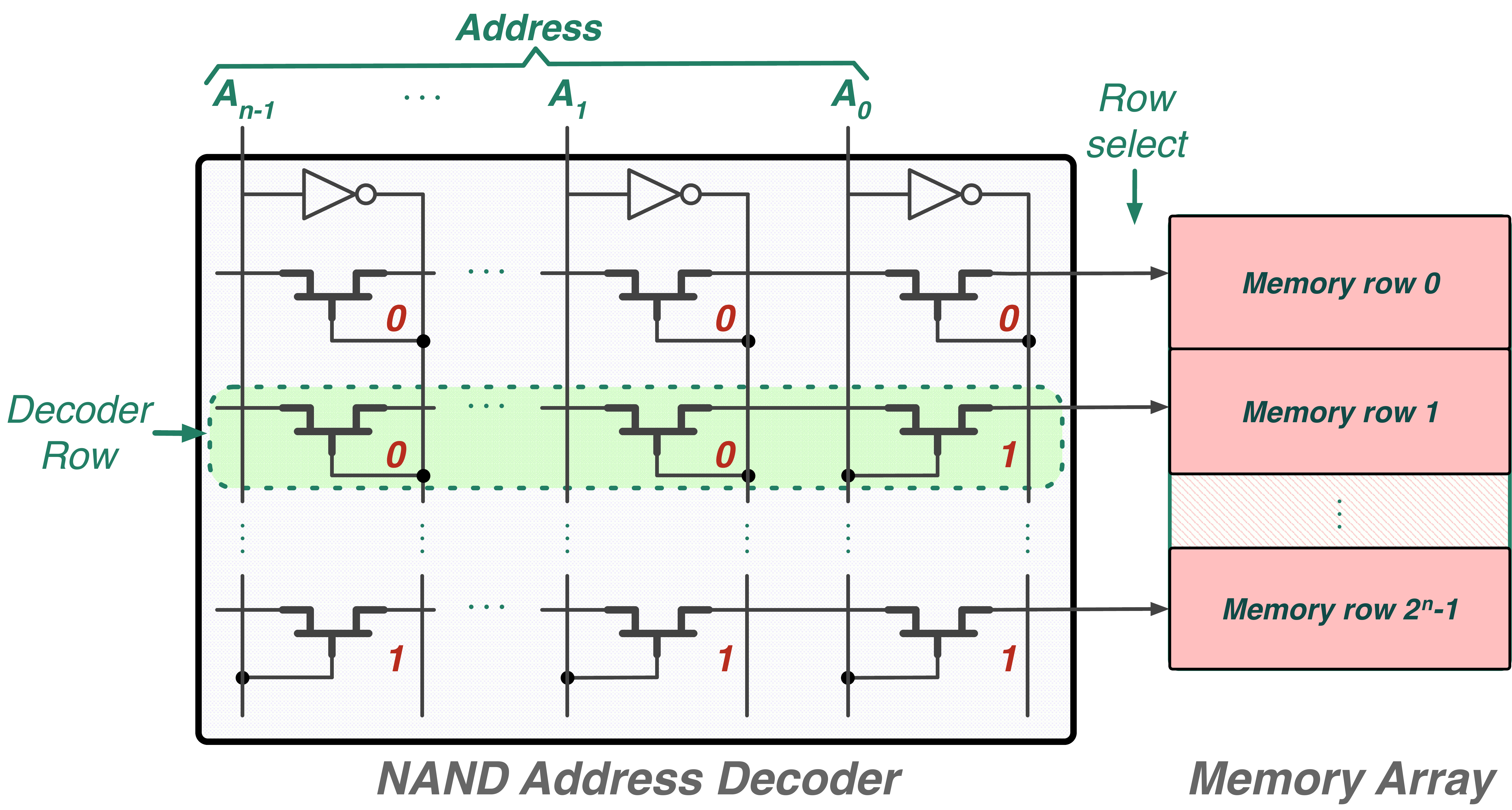}
    \caption{Conventional NAND address decoder. }
    \label{fig:conventional_decoder}
\end{figure}

Figure~\ref{fig:conventional_decoder} shows a conventional hardwired dynamic NAND address decoder that consists of an array of NMOS transistors. A conventional decoder selects a specific row of the memory array according to the input address. The mapping between the input address and the selected row in the \lois{memory} array \onur{cannot} be changed.
The gates of the NMOS transistors in each decoder row are hardwired to either direct (\joaotwo{e.g.,} $\text{A}_0$) or inverse address bitlines (e.g., $\overline{\text{A}}_0$), according to the physical position of the decoder and memory rows. Additionally, a dynamic NAND address decoder typically includes precharge transistors, evaluation \onurthree{transistors,} and a level keeper in each decoder row (not shown in \onur{Figure~\ref{fig:conventional_decoder}})~\cite{huh201264gb}.


\subsection{Programmable Resistive Address Decoder}
\label{sec:PRAD}

Our proposal relies on Programmable Resistive Address Decoders (PRADs)~\cite{yavits2017resistive} to implement both wear-leveling and fault \onur{tolerance} mechanisms for resistive memories. 

\begin{figure}[h] 
\centering
    \includegraphics[width=1\linewidth]{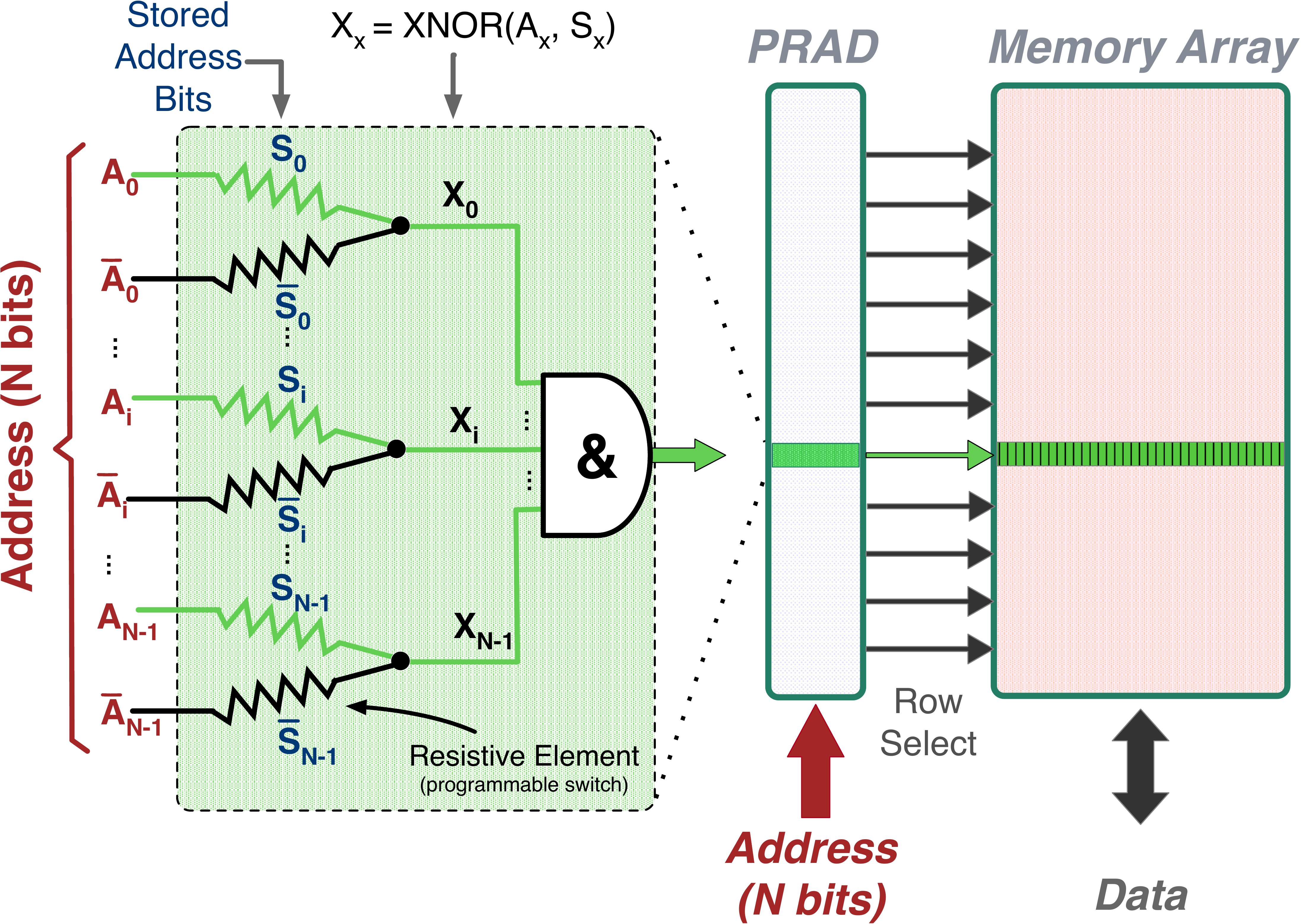}

    \caption{Programmable resistive address decoder (PRAD).}
   \label{fig:PRAD}
\end{figure}

\onur{Figure~\ref{fig:PRAD} shows a high-level overview of PRAD.} PRAD decouples memory addresses from fixed physical locations within the memory \lois{array}. \lois{PRAD} provides a level of indirection \lois{that allows} flexible and dynamic mapping of memory addresses onto arbitrary memory positions. 

PRAD allows programming addresses into address decoder positions (i.e., \lois{decoder} rows) using resistive elements. 
Each \onur{stored} address bit in a decoder row can be programmed with two resistive elements \onur{(e.g., $S_i$ and $\overline{S}_i$ in Figure~\ref{fig:PRAD})}.
During memory access, the address is looked up in a fully-associative fashion: a pair of resistive elements functions as \onur{an} XNOR gate \onur{that compares} a bit of the input address \onur{(e.g., $A_i$)} to the \onur{stored address bit (e.g., $S_i$)}. \onur{If $A_i$ and $S_i$ have the same value}, \onur{the input \onur{$i$} is asserted in} the AND (\&) gate. If all input address bits match the stored \onur{address} bits in a decoder row, the AND gate outputs ‘1’, selecting the memory row. 

\section{\mechanism{}: New Wear-Leveling \\ and Fault Tolerance Mechanisms}
\label{sec:Idea}

\mechanism{} is a new \lois{mechanism} for improving the lifetime of \onurtwo{resistive} memories that \onurtwo{seamlessly} \lois{integrates} wear-leveling and fault-tolerance at low cost. \mechanism{} is the first work \onurtwo{that combines} both techniques efficiently, achieving better memory lifetime than state-of-the-art works.

\vspace{5pt}\noindent\textbf{Hardware Components.} \mechanism{} requires three key hardware components to enable an efficient implementation of the wear-leveling and \onur{fault-tolerance} \lois{mechanisms}. First, a \lois{programmable} address decoder (PRAD) that enables efficient remapping of memory addresses via PRAD programming. PRADs  (Section~\ref{sec:PRAD}) \onur{replace} the conventional decoders (Section~\ref{sec:conv_dec}) used in common resistive memories. Second, a swap buffer (SB) that enables efficient swapping of the \onur{contents} of two memory addresses. The SB is connected to the sense amplifier in parallel with the row buffer by using multiplexers. This is possible because sense amplifiers and row buffers are decoupled in \onurfive{non-volatile} \onurfour{memories~\cite{standard2011low,meza2012case,meza2012evaluating,yoon2014efficient,Lee2009,lee2010phasescale}}. In our evaluation (Section~\ref{sec:area_overhead}), we show that the SB incurs very low hardware overhead. 
%
\onurthree{Third, a \mechanism{} controller per memory rank, placed in the memory module, that can issue memory commands to each bank independently. The  goal of having the \mechanism{} controller in the memory module is to keep the memory bus free from additional traffic generated by the wear-leveling and fault tolerance mechanisms. 
\mechanism{} controller uses the same existing commands used by the memory controller, but the PRE and ACT commands use the SB instead of the RB. 
We describe the \mechanism{} controller in detail in Section~\ref{sec:wolfram_controller}.}

\subsection{\mechanism{} Wear-Leveling}
\label{sec:WL-remap-swap}

Wear-leveling is a technique that evenly distributes write accesses across the entire \onur{memory} with the goal of wearing out all memory positions at the same pace. \mechanism{} introduces a new wear-leveling technique that improves the state-of-the-art mechanisms in two ways. First, \mechanism{} reduces hardware cost by re-using the PRAD hardware that is also used for the \mechanism{} fault tolerance mechanism. Second, \mechanism{} provides fast and effective wear-leveling by remapping memory on write accesses in a \onur{\emph{pattern-aware}} manner. 

\vspace{5pt}\noindent\textbf{Limitations of Previous Works.}
In state-of-the-art wear-leveling \lois{mechanisms}~\cite{qureshi2009enhancing,seong2010security,zhou2009durable}
, memory addresses are remapped one by one, in a sequential fashion, regardless of the actual write patterns. As a result, \lois{both} rarely and frequently written addresses are remapped at the same pace, leading to \lois{sub-optimal} write non-uniformity removal.

To avoid this problem, \mechanism{} remaps and swaps \lois{memory positions that are accessed for writing}, \onurtwo{and} thus the probability of an address to be remapped grows with its \onur{write} access frequency. In other words, \mechanism{} remaps the frequently written addresses more frequently, \onurthree{thereby} flattening the \onur{wear distribution across the entire memory at a} much faster \onur{pace than prior wear-leveling mechanisms}.

\subsubsection{Remapping and Swapping Operation}
\label{sec:remap_and_swap}
\mechanism{} performs wear-leveling by remapping pairs of addresses and swapping their \onur{contents}. \mechanism{} can perform these operations \lois{1) at fine granularity,} \onur{i.e.,} between two \lois{memory} blocks within a subarray, \lois{or 2) at \onur{course} granularity}, \onur{i.e.,} between two entire subarrays in a bank. At each write access, \mechanism{}  remaps the write address to a random location and it swaps the contents of \onur{the original and the random location}. To reduce the \onur{energy} and performance cost \onur{of the remap and swap operation}, \mechanism{} does \onur{\emph{not}} perform this operation on every write access, but with a probability \lois{such that the \onur{wear distribution is} almost the same as the \onur{wear distribution} of remap and swap on every write access}.

\begin{figure}[t] 
	\centering
    \includegraphics[width=1.0\linewidth]{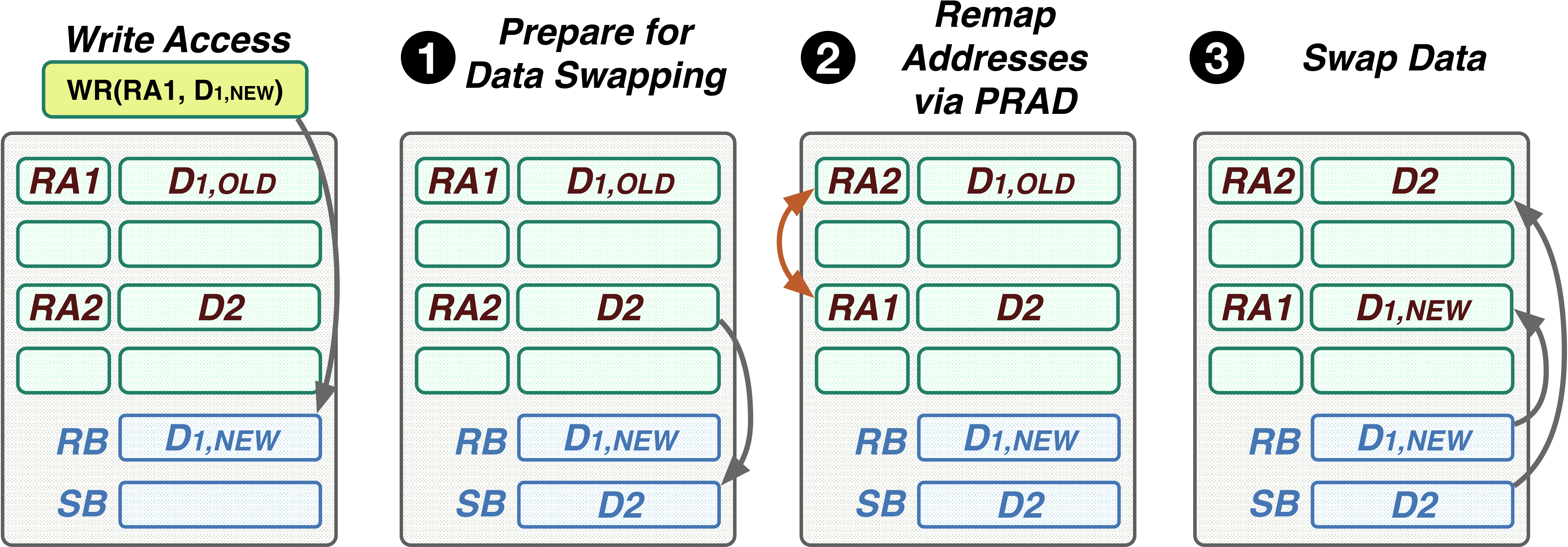}
    \caption{Example of \mechanism{} remapping and swapping two memory \onurtwo{blocks.}}
    \label{fig:remmaping}
\end{figure}

\vspace{5pt}\noindent\textbf{Remapping and Swapping \lois{Memory} Blocks.}
Figure~\ref{fig:remmaping} shows an example of remapping and swapping of two \lois{memory} blocks within a subarray. A write access to address RA1 stores its \onur{new data} (D1,NEW) into the row buffer (RB). If the write access is eligible for remapping and swapping (see Section~\ref{sec:frequency}), \mechanism{} executes three steps. First, \mechanism{} \lois{selects} a random swapping address \joaotwo{(RA2) and} copies its content to the swap buffer (SB) \circledblack{1}. Second, \mechanism{} remaps address RA1 to RA2, and RA2 to RA1 by reprogramming the PRAD \circledblack{2}. During this step, the bank becomes unavailable. Third, \mechanism{} effectively swaps the data by copying back the content of RB and SB to their original addresses \circledblack{3}. \lois{At the end of the three steps}, the two \lois{memory} blocks effectively switch their physical positions while maintaining their addresses.



\begin{figure}[ht] 
    \centering
    \includegraphics[width=1.0\linewidth]{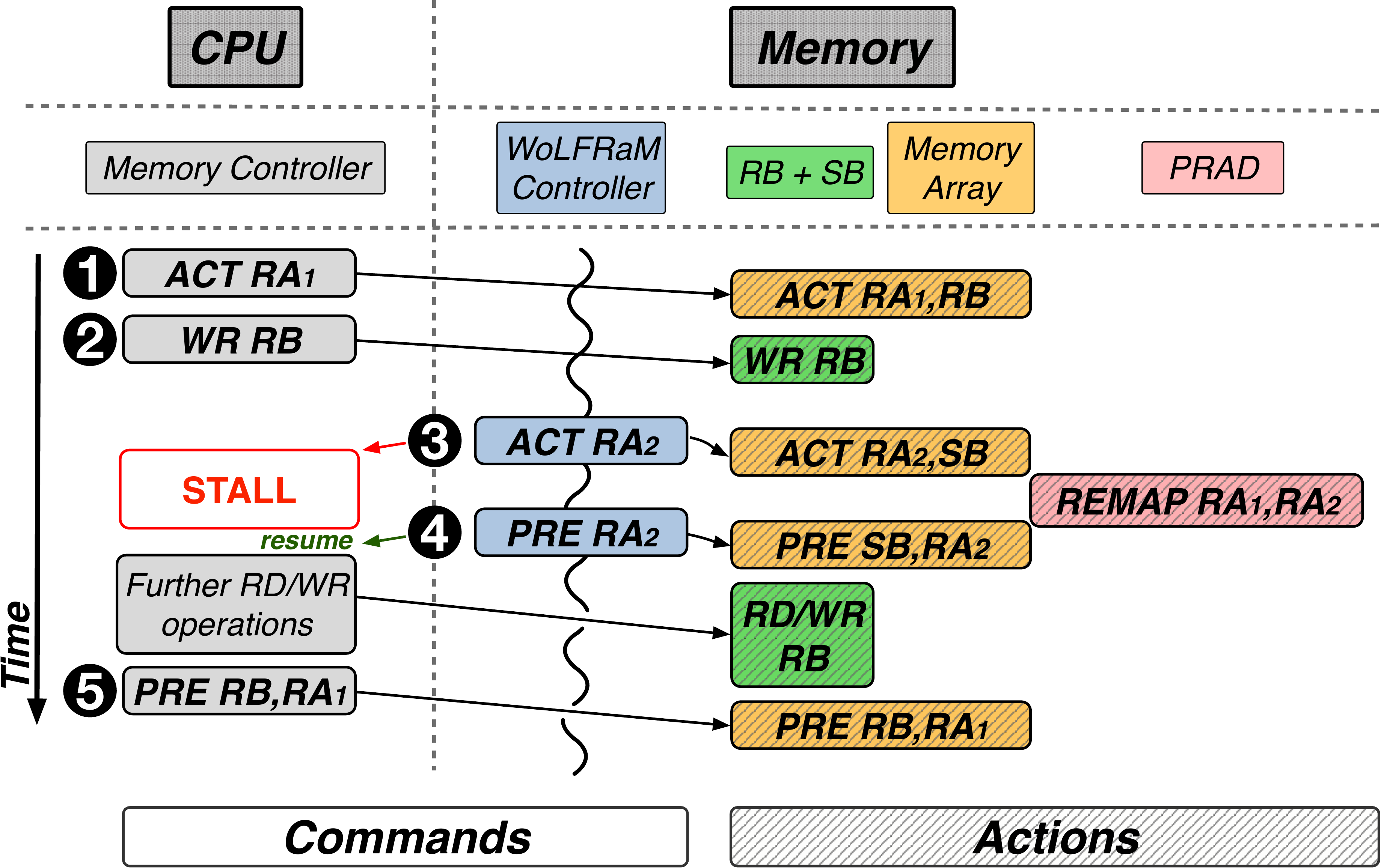}
    \caption{Sequence of commands issued by the memory controller and the \mechanism{} controller to remap and swap two memory \lois{blocks}.
    }
    \label{fig:commands}
\end{figure}

\onurthree{Figure~\ref{fig:commands} shows the sequence of commands required by the \mechanism{} controller to remap and swap a memory block after a write access from the CPU. We explain the process with a five-step example.}

First, the memory controller issues an ACT command that reads \onur{block} $\text{RA}_1$ from the memory array into the row buffer (RB) \circledblack{1}. Second, the memory controller issues a WR command that writes the \onur{new} data into the RB \circledblack{2}. Third, the \mechanism{} controller detects the \lois{WR} command from the CPU and starts the remap and swap operation by selecting a random block ($\text{RA}_2$) and issuing an ACT command that brings the content of $\text{RA}_2$ into the swap buffer (SB) \circledblack{3}. This step also executes the key operation of reprogramming the PRAD to switch $\text{RA}_2$ and $\text{RA}_1$ addresses. \onurthree{To avoid conflicting commands from the memory controller, the \mechanism{} controller signals the memory controller to \onurfive{stop} issuing commands to the memory bank \onursix{(STALL)} while the remap and swap operation is executing.} Fourth, the \mechanism{} controller issues a PRE command that writes back the content of SB  into its original address $\text{RA}_2$ \circledblack{4}, which is now placed where $\text{RA}_1$ was placed before the remapping \lois{operation}. \onurthree{\onurfour{\onurfive{As the} \mechanism{} controller completes} the PRE command, \onurfive{it} sends a \onurfive{resume} signal to the memory controller\onurfive{, indicating that it can issue} \onurfour{commands} to the memory bank again.} Fifth, the memory controller issues a PRE command that writes back the content of \onur{RB} into its original address $\text{RA}_1$ \circledblack{5}, which is now placed where $\text{RA}_2$ was placed before the remapping \lois{operation}. At the end of this process, the two blocks are effectively swapped in the physical space. Note that the memory controller can issue regular RD/WR commands freely if the PRAD is not being reprogrammed.

\vspace{5pt}\noindent\textbf{Remapping and Swapping Subarrays.}
When a subarray receives many write accesses, \mechanism{} might decide to remap and swap the entire subarray (see Section~\ref{sec:frequency}). The process \joaotwo{consists} of two main steps.
First, \mechanism{} selects a random subarray to perform the remap and swap \lois{operation}.
Second, \mechanism{} controller issues remap and swap commands to \emph{all} blocks in the subarray. Because all subarrays in a bank \onur{\emph{share}} the row buffer~\cite{standard2011low,meza2012case,meza2012evaluating,yoon2014efficient,Lee2009,lee2010phasescale}  and the swap buffer, the remap and swap operation of each individual block \onurtwo{is similar} to \lois{the} remap and swap \lois{operation} within a subarray. The difference is that for remapping and swapping a subarray, \lois{\mechanism{} reprograms} the global PRAD instead of the local PRAD (see Figure~\ref{fig:memory_bank}).

\subsubsection{Remapping and Swapping Frequency}
\label{sec:frequency}

To limit the performance impact of the remap and swap \onur{operations} and additional PRAD wear caused by extra \lois{programming operations}, \mechanism{} remaps \lois{and swaps} \onurtwo{at} a sufficiently low \onurtwo{frequency}. \joao{The} \mechanism{} controller implements this mechanism by generating a random number \onur{(e.g., via a mechanism similar to D-RaNGe~\cite{kim2019d})} on every write access. If the generated number is less than or equal \onurtwo{to threshold} $\sigma_1$, \mechanism{} \lois{remaps and swaps the write address within the subarray}, and if it is less than or equal to threshold $\sigma_2$, \mechanism{} remaps and swaps the entire subarray. \onur{The higher the $\sigma_1$ and $\sigma_2$ thresholds, the faster the \joaotwo{wear-leveling}, at the cost of higher performance and energy overheads. In our evaluation, the $\sigma_2$ \joaotwo{threshold} is much lower than $\sigma_1$, as remapping an entire subarray is much \onurtwo{more} costly than remapping a single \onurtwo{memory block} (e.g., 512$\times$} energy and performance \onurthree{overhead}). 
Previous works propose similar \onurtwo{randomized} swapping techniques in the context \lois{of} \onur{wear-leveling} for flash memories~\cite{ben2006competitive} and PCM as secure main memory~\cite{Seznec2010}. Our evaluation (Section~\ref{sec:evalperformance_energy})  shows that \mechanism{}  performance overhead is very low.

\subsubsection{Preventing \onur{Wear-Out} Attacks}
\label{sec:random_remap_period}

\mechanism{} is secure against attacks that try to wear out a particular memory position. The probabilistic approach implemented in \mechanism{} renders such \onur{an} attack impractical, since the remapping intervals are entirely random. We quantitatively demonstrate this in Section \ref{sec:eval_lifetime} for \onur{the} repeated address attack.

Unlike \mechanism{}, simple wear-leveling approaches~\cite{seong2010security,qureshi2009enhancing} 
use a constant remapping interval that \onur{triggers} \onurtwo{subarray-level} remapping \onurtwo{exactly every} $n_{th}$ write access. Such approaches create an opportunity for malicious \lois{exploits}~\cite{Seznec2010}: after inferring the remapping interval size $n$, the attacker \lois{may} wear a certain memory position by writing to it $n-1$ times, and changing the address on the $n_{th}$ write, so the mechanism remaps an unrelated address. The attacker can repeat this procedure during every remapping cycle, which can significantly reduce the efficiency of the wear-leveling \lois{mechanism}. 

\subsection{\mechanism{} Fault Tolerance}
\label{sec:wolfram_faultrecovery}

\mechanism{} fault tolerance mechanism can recover from a memory block failure by remapping such \onur{a} failed memory block \onurtwo{to} an empty (unoccupied or spare) memory \onur{block}. \mechanism{} tracks empty \lois{memory} \onur{blocks} in hardware \lois{using} an additional bit that is set when a new address is programmed into the PRAD, and reset when a data block is deleted.

\mechanism{} detects failures via read-after-write verification~\cite{qureshi2009enhancing}. In \mechanism{}, a memory block failure does not \lois{require} mapping-out an entire page as proposed \onur{by} \lois{many} resistive memory fault \onur{tolerance} \lois{mechanisms}~\cite{schechter2010use,fan2013aegis,qureshi2011pay,azevedo2013zombie,ipek2010dynamically,seong2010safer}. \mechanism{} enables \lois{fine-grained remapping at memory block \onur{granularity}), which allows the memory to continue operating transparently while its capacity reduces with the number of memory block failures.} 

A failed \lois{memory} block is remapped by (1) physically disabling the PRAD row associated with it, and (2) reprogramming its address into an empty \lois{PRAD row}. \lois{As a result,} the address of the \lois{memory} block remains the same although its physical location changes.

\begin{figure}[h] 
\centering
    \includegraphics[width=0.9\linewidth]{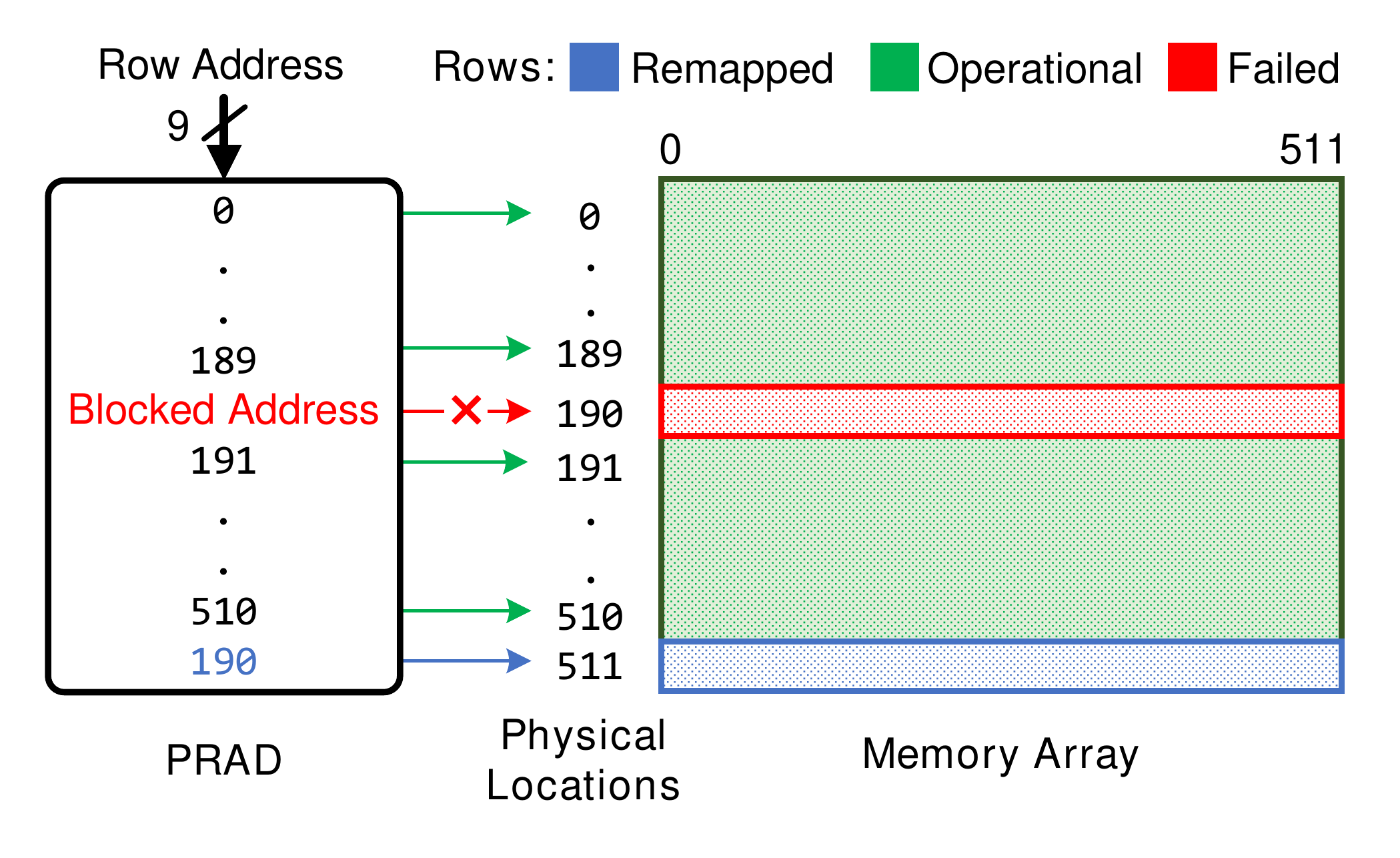}
    \caption{Example of \lois{\mechanism{}} block failure recovery.}
    \label{fig:block_recovery}
\end{figure}

\vspace{5pt}\noindent\textbf{Block failures.}
Figure~\ref{fig:block_recovery} illustrates how \mechanism{} repairs a memory block failure within a subarray where all addresses are originally programmed to match \lois{their} physical locations. When \mechanism{} detects a failed \lois{memory} block, it remaps the failed \lois{memory} block to an empty row in the subarray. In the example of the \onur{Figure~\ref{fig:block_recovery}}, \mechanism{} programs the address of the failed block $190$ into the empty row $511$ at the \onur{bottom} of the subarray. \onurtwo{After that point, the physical address $190$ is marked as blocked and never accessed again (i.e., it is mapped out of the address space).}

\vspace{5pt}\noindent\textbf{Subarray failures.}
If a subarray experiences a terminal failure (\lois{e.g.,} most of its memory blocks fail), it can be remapped \onurtwo{to} an empty subarray by reprogramming the global PRAD.

\subsection{Combining \mechanism{} with Existing Fault Correction Techniques}
\label{sec:fault_correction}


\mechanism{} is compatible with many state-of-the-art fault correction techniques, which allows tolerating more than one fault per memory block.  

We briefly discuss two fault correction techniques that can be easily \onur{integrated} with \mechanism{}.
First, \mechanism{} can be combined with \onur{ECP~\cite{schechter2010use}} by replacing \onur{the} hardwired address decoder of the memory device \onurtwo{with} PRAD. ECP stores several error correcting pointers in each memory block and replaces failed cells with redundant ones.
Unlike \lois{the original ECP paper~\cite{schechter2010use}}, \mechanism{} does not \lois{require} \onurtwo{recovering} a terminal \lois{memory} block failure by decommissioning the entire page. Instead, \mechanism{} \onur{simply} remaps the failed \lois{memory} block \onur{to a known-operational memory location.}

Second, \mechanism{} can be integrated with Error Correcting Codes \onurfive{(ECC)~\cite{hamming1950error}}. Similar to FREE-p~\cite{yoon2011free}, \mechanism{} can be integrated with several ECC schemes, including simple ECC schemes and chipkill~\cite{luo2014characterizing}.

\subsection{Putting it \onur{All Together}:  Wear-Leveling + Fault Tolerance}
\label{sec:wolfram_integration}

When a memory block fails and is mapped out (i.e., \onur{its} address is removed from the address space), the wear-leveling mechanism \lois{should} \onurtwo{no} longer use this address for remapping.
\mechanism{} resolves this issue by simply OR-ing all \lois{row-selects} in \lois{the} PRAD. \lois{The OR Output '0' indicates} that there is no matching decoder position (\lois{i.e.,} the looked-up address belongs to a mapped-out block), \lois{so} the wear-leveling controller reattempts the remapping. Since no actual write is made into a mapped-out location, the performance overhead of \onur{\mechanism{}'s remapping \onurtwo{attempt} is negligible.}

State-of-the-art fault tolerance techniques~\cite{schechter2010use,seong2010safer,fan2013aegis,yoon2011free,qureshi2011pay,azevedo2013zombie,ipek2010dynamically} do not discuss how wear-leveling can continue operating seamlessly after a \onur{failed} memory block is mapped out. Once a block fails, the assumption that any address can be remapped to any other address is no longer valid~\cite{Fan2014}. 
\lois{One way to solve this problem is \onur{to} detect mapped-out locations by checking failures in the read-after-write verification process. If the verification fails, the wear-leveling mechanism should reattempt the remapping and writing. This approach incurs additional performance overhead \onurtwo{due to} the additional write 
operations.
} 

\subsection{\onurthree{\mechanism{} Controller}}

\label{sec:wolfram_controller}

\onurthree{There are several ways to implement the \mechanism{} controller in a resistive memory system. We use a \mechanism{} controller per memory bank, and we \onurfour{place all \mechanism{} controllers} in one separate chip in the memory module \onurseven{(similar to~\cite{seshadri2015gather})}. Each \mechanism{} controller can issue memory commands to \onurfour{its associated} memory bank, and its operation is independent of the \onurfour{other} \mechanism{} controllers \onurfour{for different banks}. We find two main challenges in implementing the \mechanism{} controller.}

\onurthree{First, every time the \mechanism{} controller executes a remap and swap operation, it needs to notify to the memory controller that it should not issue any command to the memory bank while the swap and remap operation is executing. To enable the synchronization between the  \mechanism{} controller and the memory controller, we add a new pin in the DRAM module. \onurfive{Before the remap and swap operation starts, the \mechanism{} controller sends} a synchronization signal \onurfive{on this pin to indicate that} commands from the memory controller to the bank \onurfive{should stall}. \onurfour{When the remap and swap operation finishes, the \mechanism{} controller sends a synchronization signal \onurfive{on the same pin to indicate that commands from the memory controller to the bank can resume}.}}

\onurthree{Second, \mechanism{} uses a a probabilistic approach to remap and swap memory blocks (Section~\ref{sec:frequency}), which \onurfour{requires} generating random numbers. We use a true random number generator \onurfive{(TRNG),} called D-RaNGe~\cite{kim2019d}, that reduces the memory access latency below reliable values and \onurfour{exploits} memory cells' failure probability to generate random numbers. \mechanism{} controller 1) generates random numbers when the chip is idle, 2) compares the generated random values to $\sigma_1$ and $\sigma_2$ thresholds to decide if \onurfour{it} needs to remap and swap future write accesses, and 3) it stores its decisions in a small array of bits. An alternative implementation is to use a \onurfour{pseudo-random} number generator (PRNG)~\cite{von195113}, which uses \onurfour{a deterministic} algorithm to generate a sequence of random numbers from a seed value. A PRNG \onurfour{avoids adding a new pin to the memory module for synchronization:} we can synchronize the \mechanism{} controller and the memory controller \onurfour{by} implementing the same PRNG in both controllers, and sharing the same seed, which allows the memory controller 
to know when and for how long to stall.
\footnote{\onurfour{The drawback of using a PRNG is that a malicious attacker can reproduce the sequence of generated random numbers if they are able to obtain the PRNG seed, which could compromise the system. We choose to use a TRNG instead of a PRNG  for security reasons.} }}

\section{Experimental Setup}
\label{sec:experiemntal_setup}

We evaluate the lifetime of resistive memories by using an \onurtwo{in-house simulator}.
We compare \mechanism{} with two-level Security Refresh (SR)~\cite{seong2010security}. SR is a dynamic randomized address mapping scheme that swaps data using random keys upon each \onur{refresh}.
To ensure lifetime evaluation fairness, we select the design and simulation parameters such that \mechanism{} and SR have similar area, performance and energy overheads.  

We configure SR following the assumptions made by the original paper~\cite{seong2010security}.
For practical purposes, we select slightly suboptimal number of SR subregions (2,048 instead of the optimal SR subregion count of 1,024~\cite{seong2010security}). This allows confining the subregion to a single subarray, \lois{which} significantly \lois{reduces} the complexity of address generation.
Since SR performs two extra writes per swap~\cite{seong2010security} ($vs.$ one extra write per intra-subarray swap \lois{with \mechanism{}}\onur{, as explained in Section~\ref{sec:remap_and_swap}}), we apply \lois{an} inner SR refresh interval of 200 \lois{write accesses} to ensure a fair comparison. The outer SR refresh interval is set \lois{to} 100 \lois{write accesses}, on par with the average \mechanism{} inter-subarray remapping interval.

We configure \mechanism{} \lois{for remapping} individual memory blocks with $\sigma_1$=1\% probability (i.e., the average remapping interval is 100 write accesses),
because it provides a good trade-off between performance overhead and wear-leveling.
We choose to remap \onur{an} entire subarray with a probability $\sigma_2$=0.002\% \onurtwo{(i.e., the average remapping interval is 512x100 write accesses)} such that the performance overhead is similar to that of individual memory block \lois{remapping}.

We calculate the area of the swap buffer (SB) used in our evaluation by using data from prior work~\cite{Lee2009}.
We evaluate the energy and latency of PRAD using Cadence Virtuoso~\cite{Virtuoso} with a 28nm high-K metal gate library from \onur{GlobalFoundries}. We verify the functionality of PRAD, and simulate its timing and energy consumption using SPICE \onur{simulations}~\cite{Nagel1973}.

Table~\ref{table:PRAD} shows the latency and energy of the \onur{baseline} 9-to-512 NAND hardwired address decoder and the 9-to-512 NAND PRAD we use in our evaluation. We also show the overhead of PRAD compared to \lois{a} hardwired address decoder, and compared to the entire memory subarray.

\begin{table}[h]
    \centering
    \footnotesize
    \setlength\tabcolsep{4pt} 
    \begin{tabular}{c?{3\arrayrulewidth}m{1.3cm}|m{1.3cm}|m{1.3cm}|m{1.3cm}}
     & {\bf Hardwired Address Decoder}  & {\bf PRAD} & {\bf PRAD/
     Hardwired Decoder Overhead} & {\bf PRAD/Memory Subarray Overhead}\\ \Xhline{3\arrayrulewidth}
    {\bf Latency} & 112.2 ps & 112.7 ps & 0.44\% & 0.09\% \\
    {\bf Energy} & 0.54 pJ & 0.63 pJ & 18\% & 0.07\%\\
    \end{tabular}
    \caption{Latency and energy of the \onur{baseline} 9-to-512 NAND hardwired address decoder and 9-to-512 NAND PRAD.}
    \label{table:PRAD}
\end{table}

We assume that a memory cell lifetime (\lois{i.e.,} write endurance) is normally distributed with the mean of $10^8$ writes and coefficient of variation of 15\%, \onur{similar to previous works~\cite{seong2010security,qureshi2009enhancing,qureshi2011pay,yoon2011free,qureshi2011practical}.}

\vspace{5pt}\noindent\textbf{Performance.}
\lois{To evaluate \onurfour{performance,} we use the \onur{state-of-the-art extensible} DRAM simulator Ramulator~\cite{kim2015ramulator,ramulatorgithub} extended to support PRAD.
We open-source our simulator and all configurations used for collecting our results~\cite{ramulatorWolfram}.
\onurtwo{To collect} the Ramulator input memory traces, we use Intel’s \onur{dynamic} binary instrumentation tool, Pin~\cite{Luk2005}, for all the benchmarks described in Section~\ref{sec:workloads}.}

Table~\ref{table:memory_config} \lois{shows} the configuration of the PCM memory system. Each 1GB bank has \onurtwo{a row size} of 1KB and consists of $2^{20}$  rows~\cite{seong2010security}. The interface \lois{used by the} memory controller is LPDDR2-NVM-based~\cite{LPDDR2-NVM}, where each read/write is a burst of eight 64b transfers \lois{(i.e., 64B per request)}~\cite{lee2015decoupled,Kim2012,lee2013tiered}. 

\begin{table}[ht]
    \centering
    \footnotesize
    \begin{tabular}{r?{3\arrayrulewidth}l}
    \hline
    {\bf Memory Type} & Phase Change Memory (PCM) \\ \hline
    {\bf Banks} & 1 GB capacity, 1KB ($2^{13}$ bits) row \onurtwo{size}, $2^{20}$ rows  \\ \hline
    {\bf I/O} & \onurtwo{\begin{tabular}[x]{@{}l@{}}400 MHz, 800 MT/s max transfer rate, \\ 8 burst length, 64b channel width \end{tabular}} \\ \hline
    \end{tabular}
    \caption{\lois{Resistive main} memory configuration.}
    \label{table:memory_config}
\end{table}

Table~\ref{table:energy_timing} shows the timing and energy parameters of the PCM memory \lois{used as main memory in our evaluation}. To ensure \lois{a} fair comparison with Security Refresh, the timing and energy values are based on the data provided in~\cite{Lee2009}.

\begin{table}[ht]
    \centering
    \footnotesize
    \begin{tabular}{|c|c|}
    \hline
    {\bf \lois{Operation}} & {\bf Energy (pJ/bit)} \\ \hline
    Array read & 2.47 \\
    Array write & 16.82 \\
    Buffer read & 0.93 \\
    Buffer write & 1.03 \\
    Standby & 0.08 \\
    \hline
    \hline
    {\bf \lois{Timing parameters}} & {\bf \lois{Cycles}} \\ 
    \hline
    tRCD, tCL, tWL, tCCD, tWTR & 22, 5, 4, 4, 3 \\
    tWR, \onur{tRTP}, tRP, tRRDact, tRRDpre & 6, 3, 60, 2, 11 \\
    \hline
    \end{tabular}
    \caption{Energy and timing parameters of the \lois{evaluated PCM} main memory system.}
    \label{table:energy_timing}
\end{table}

Table~\ref{table:cpu_configuration} shows the CPU configuration used in our Ramulator simulations. 
\lois{We \onurfour{obtain} results by running} each simulation for 1.4 billion instructions, \lois{after 1 million warmup instructions}.

\begin{table}[ht]
    \centering
    \footnotesize
    \setlength\tabcolsep{3pt}
    \begin{tabular}{r?{3\arrayrulewidth}m{5.4cm}}
    \hline
    {\bf \onurfive{Processor}} & \onurfive{Single-core}, 2.4GHz, \onurfive{L1-I 32kiB, L1-D 32kiB, ~~~~~L2 2MiB, L3 16MiB} \\ \hline 
    {\bf Memory controller} & \multirow{ 2}{*}{32} \\
    {\bf R/W Queue Size} & \\\hline
    {\bf Memory Scheduler} & FRFCFS with cap of 16 on row \onurfour{hits~\cite{Mutlu2007,subramanian2014blacklisting,subramanian2016bliss,mutlu2008parallelism}} \\\hline
    {\bf Address translation } & MSB to LSB: Row $\rightarrow$ Bank $\rightarrow$ Rank $\rightarrow$ Column $\rightarrow$ Channel \\\hline
    \end{tabular}
    \caption{CPU configuration.}
    \label{table:cpu_configuration}
\end{table}


\vspace{5pt}\noindent\textbf{Energy.} To evaluate the \lois{\mechanism{}} energy \onurtwo{consumption,} we use an in-house energy estimation tool connected to Ramulator that estimates the read/write energy based on parameters \onur{from~\cite{Lee2009}, summarized} \onurtwo{in} Table~\ref{table:PRAD} and Table~\ref{table:energy_timing}.

\subsection{Workloads}
\label{sec:workloads}
\vspace{5pt}\noindent\textbf{Attack test.} For memory lifetime evaluation, we use a \emph{repeated address attack test} that repeatedly writes to the same memory location~\cite{seong2010security}. This test is the simplest malicious wear-out attack.

\vspace{5pt}\noindent\textbf{\onur{SPEC CPU2006}.} For performance and energy overhead evaluation, we \lois{use 28 benchmarks from}  \onurseven{SPEC CPU2006~\cite{Henning2007}}. 

\section{Evaluation}
\label{sec:evaluation}

\subsection{Area Overhead}
\label{sec:area_overhead}

Table~\ref{table:area_overhead} shows the area overhead of the evaluated mechanisms relative to the size of a \lois{PCM memory} subarray. \onur{At} the top \onur{part} of the table we show the \onur{overheads} of the evaluated mechanisms alone \lois{(SR, \mechanism{})}.
\onur{At} the bottom \onur{part} of the table we show the \onur{overheads of the} \lois{same mechanisms} \onur{when} combined with ECP error correction~\cite{schechter2010use}. 
\onurtwo{An} $\text{ECP}_k$ \onur{error} correction \lois{mechanism can correct} k failed bits in an individual memory block \lois{with} an area overhead of approximately \lois{$k\times1.90$\%}~\cite{schechter2010use}.

\begin{table}[ht]
    \centering
    \footnotesize
    \setlength\tabcolsep{3pt} 
    \begin{tabular}{r?{3\arrayrulewidth}c}
    {\bf Mechanism} & {\bf Area Overhead \onur{(in Subarray)}} \\ \Xhline{3\arrayrulewidth}
    SR & 0.02\% \\
    \mechanism{} & 0.95\%\\ 
    \hline
    $\text{SR}+\text{ECP}_1$ & \lois{1.90\%} \\ 
    $\text{SR}+\text{ECP}_7$ & 13.33\%  \\ 
    $\text{\mechanism{}}+\text{ECP}_6$ & 12.85\% \\ 
    \end{tabular}
    \caption{Area overhead of the evaluated mechanisms.}
    \label{table:area_overhead}
\end{table}

Our results show that the area overhead \lois{of \mechanism{}} is slightly below 0.95\%, \onur{of} which 0.85\% \onurtwo{is from} PRADs, and 0.1\% \onurtwo{is from} the rest of the hardware components. Although \lois{the area overhead of \mechanism{} is higher than that of SR}, \mechanism{} \lois{provides better protection against errors (i.e., both wear-leveling and fault tolerance \onur{as we show in Section~\ref{sec:eval_lifetime}}).}

For approximately the same area \onur{overhead}, \mechanism{} can be combined with $\text{ECP}_6$ \lois{(6-bit correction)}, and SR can be combined with $\text{ECP}_7$ \lois{(7-bit correction)}.
\onur{In addition to ECP capabilities},
$\text{\mechanism{}}+\text{ECP}_{6}$ differs from $\text{SR}+\text{ECP}_{7}$ in that the latter maps out the entire page \lois{that contains} the failed memory block. In contrast, \mechanism{} enables decommissioning individual failed memory blocks, \lois{which} allows more graceful memory capacity degradation.

\subsection{Memory Lifetime with Wear-Leveling,\\ Fault Tolerance, and \onur{Error} Correction}
\label{sec:eval_lifetime}

To evaluate the \onur{relative effectiveness} of the wear-leveling, fault tolerance, and \onur{error} correction techniques, \lois{the metric we use} \onurtwo{is} \onur{\emph{\onurfour{usable} memory capacity as a function of the memory lifetime}}. The higher the memory capacity at any given point in the memory \lois{lifetime}, the \onur{more effective} the combined wear-leveling, fault tolerance, and \onur{error} correction techniques.

Although SR is not able to handle wear-leveling after mapping out memory pages, we assume it can continue its operation. This requires applying additional resources to make the combined SR+$\text{ECP}_k$ work~\cite{Fan2014}, \onurtwo{whose overheads} we do not account \onur{for so that we give the benefit of doubt to SR.}

\begin{figure}[ht] 
\centering
    \includegraphics[width=1.0\linewidth]{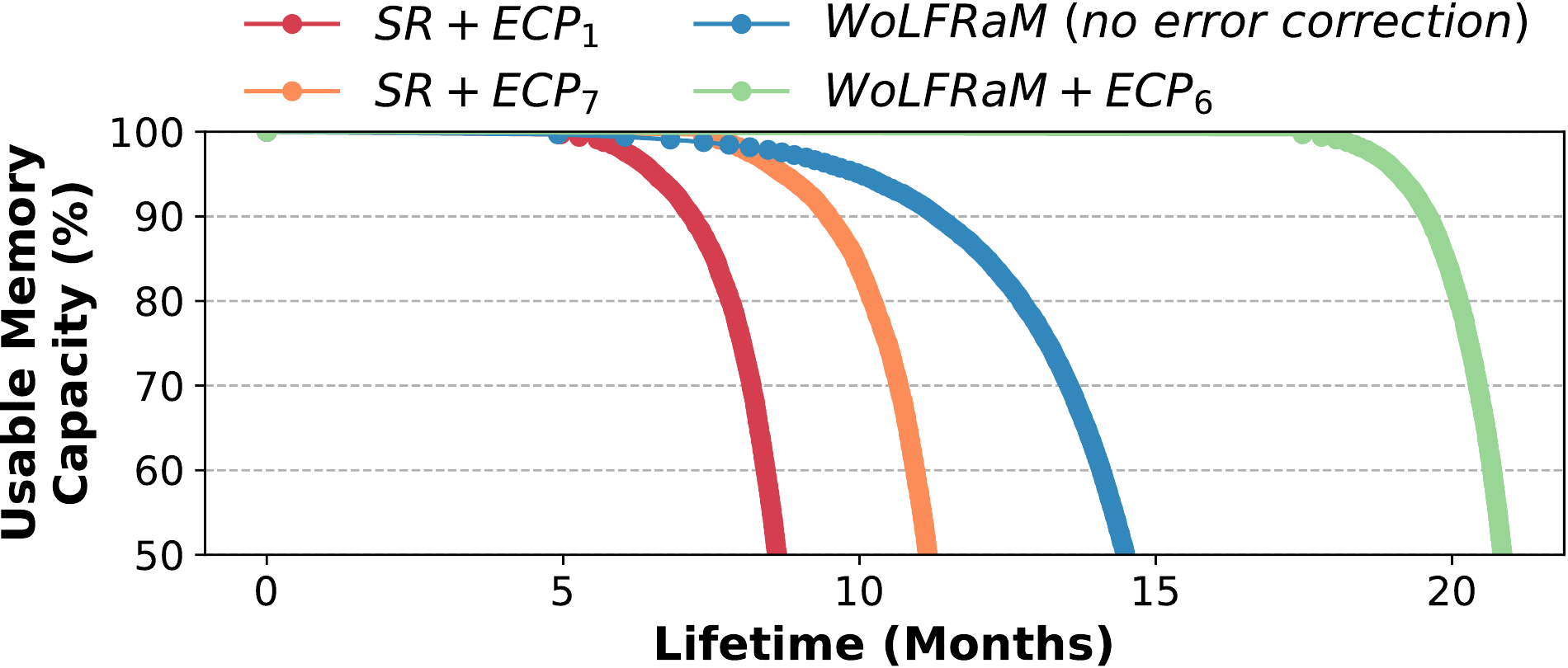}
    \caption{\onurtwo{Usable} memory capacity $vs.$ lifetime \onurfour{when running} \onurtwo{the} attack test workload.}
    \label{fig:capacityVSlifetime}
\end{figure}

Figure~\ref{fig:capacityVSlifetime} presents \onurtwo{usable} memory capacity as a function of lifetime for $\text{SR}+\text{ECP}_1$, $\text{SR}+\text{ECP}_7$, \mechanism{} with no error correction and $\text{\mechanism{}}+\text{ECP}_6$, \lois{when executing} \onurtwo{the} attack test \lois{workload} (Section~\ref{sec:workloads}). We assume that the memory device is decommissioned after its \onurtwo{usable} capacity declines by half.

\lois{We make two observations.
First,} the memory lifetime of $\text{\mechanism{}}+\text{ECP}_6$ is 87\% longer than that of state-of-the-art $\text{SR}+\text{ECP}_7$, \lois{using a similar area overhead (12.85\% $vs.$ 13.33\%).
Second, the memory lifetime of \mechanism{} with no added \onur{error} correction capabilities  is 68\% longer than that of $\text{SR}+\text{ECP}_1$, using \onur{slightly less} area overhead (0.95\% $vs.$ 1.90\%). We conclude that \mechanism{} achieves significantly longer lifetime than the state-of-the-art mechanism for similar area overhead.}  

\lois{An} additional advantage of \mechanism{} is the wear-leveling speed. A quick leveling of write non-uniformity is important and might become critical \onur{when there is} significant endurance variation \onur{across memory banks}.
In wear-leveling solutions \onur{where remapped addresses are independent of write accesses,} 
such as SR, it takes considerable time for a data block to be remapped. This means that especially "weak" cells may fail before their addresses are remapped. In contrast, \mechanism{} \onurtwo{chooses to remap} \onur{frequently-written} addresses, which \onur{allows} faster remapping of cells \onur{that have} \onurtwo{a} higher probability to fail. 

\begin{figure}[h] 
\centering
    \includegraphics[width=0.48\textwidth]{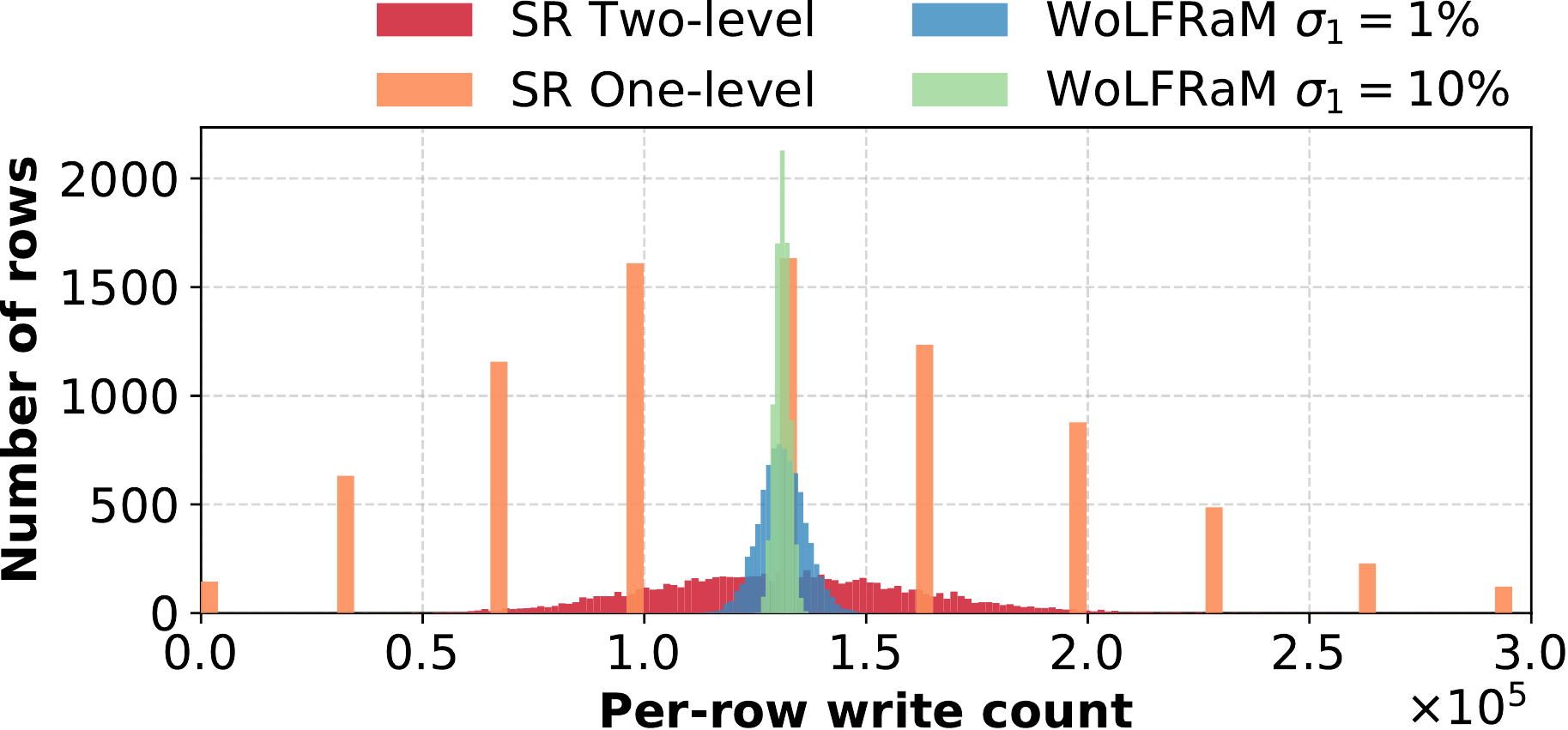}
    \caption{\onur{Per-row write count distribution.}}
    \label{fig:leveling_speed}
\end{figure}

\lois{Figure~\ref{fig:leveling_speed} shows the per-row write count histogram for \mechanism{} \onurtwo{($\sigma_1$=1\% and $\sigma_1$=10\%)}
, and single-level SR, using the attack test (Section~\ref{sec:workloads}). The ideal wear-leveling mechanism would reach \onur{an} identical number of per-row writes in each memory row, producing a single vertical bar in the per-row write count histogram. The narrower the distribution, the more \onur{effective} the wear-leveling mechanism is. We make the key observation that \mechanism{} is significantly more \onurtwo{effective} than SR, \onur{and} \mechanism{} \onurtwo{$\sigma_1$=10\%}
\onur{is} very close to the ideal wear-leveling mechanism.} 



\begin{figure}[t] 
\centering
    \includegraphics[width=0.48\textwidth]{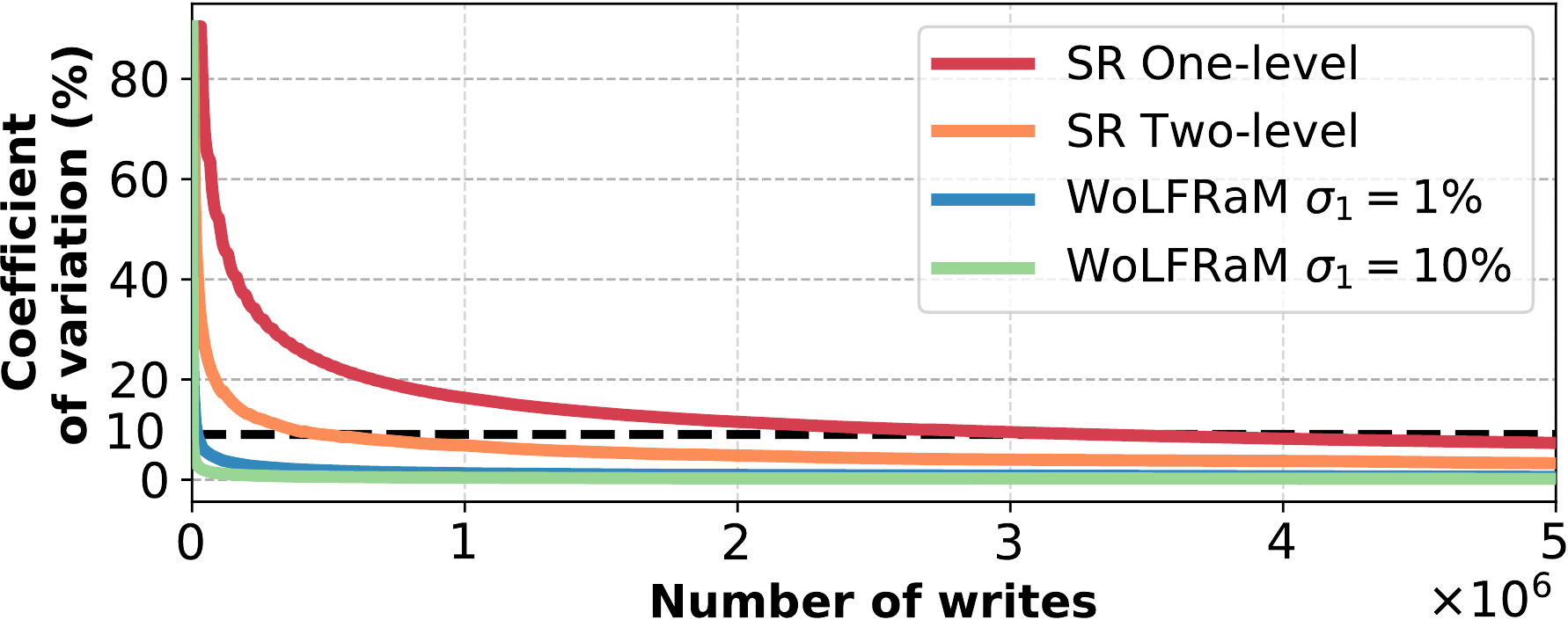}
    \caption{Coefficient of variation of the per-row write count distribution \onur{(sharper drop is better)}.}
    \label{fig:leveling_speed2}
\end{figure}

\lois{Figure~\ref{fig:leveling_speed2} shows the coefficient of variation (\onur{CoV}) of the per-row write count distribution as a function of the number of write accesses. The sharper the drop, the quicker the write nonuniformity leveling is. We make the key observation that \mechanism{} \onur{converges} much faster than SR.
\mechanism{} \onurtwo{($\sigma_1$=1\%)} 
\onur{has} 90\% \onur{CoV} drop after 21,969 write accesses, which is 21.7× faster than the two-level SR \onur{mechanism and} 147.1x faster than the one-level SR mechanism. We conclude that \mechanism{} levels the write nonuniformity significantly faster than the state-of-the-art \onur{wear-leveling} mechanisms.}

\subsection{Performance and Energy \onur{Overheads}}
\label{sec:evalperformance_energy}

\begin{figure}[ht] 
\centering
    \includegraphics[width=1.0\linewidth]{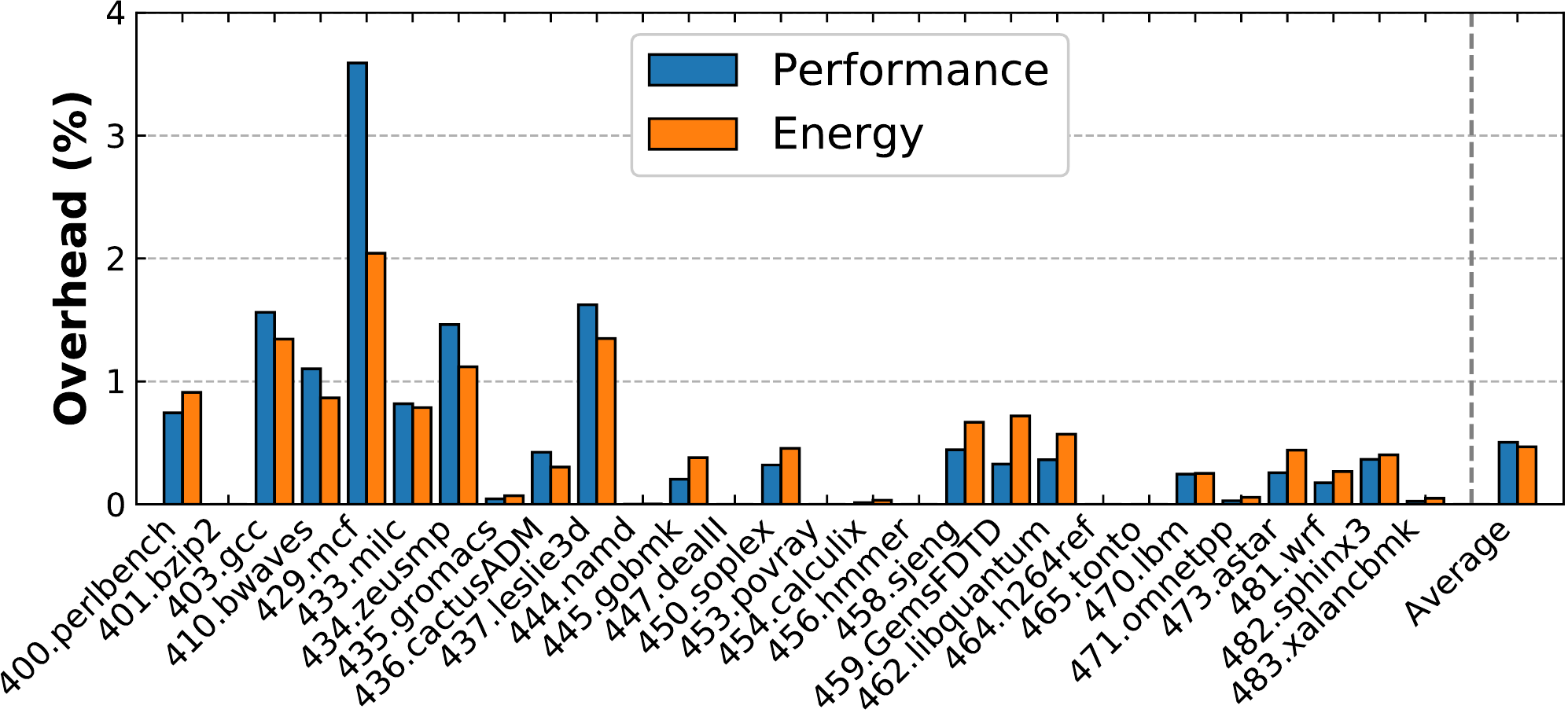}
    \caption{\mechanism{} performance and energy overheads for the SPEC CPU2006 applications for \onurtwo{$\sigma_1=1\%$ and $\sigma_2=0.002\%$}.}
    \label{fig:performance_energy}
\end{figure}

Figure~\ref{fig:performance_energy} shows the performance and energy \onur{overheads} of \mechanism{} when running SPEC CPU2006 benchmarks, with the configuration described in Section~\ref{sec:experiemntal_setup}.
We make two main observations.
First, the average performance degradation caused by \mechanism{} is only 0.51\%, and the worst performance degradation is 3.8\% (in 429.mcf). Also, there are 9 benchmarks (e.g., 444.ramd) that have negligible overhead.
Second, the average \mechanism{} energy overhead is only 0.47\%, and \onur{the worst energy overhead is only 2.1\% (429.mcf)}. Also, there are 9 benchmarks that have negligible energy overhead.

We conclude that performance and energy \onur{overheads} of \mechanism{} \onur{are} very low, and for many benchmarks the \onur{overheads are} negligible, which makes \mechanism{} a \onur{low-cost} mechanism to expand the lifetime of resistive memories.

\subsection{PRAD \onur{Wearout}}
\label{sec:PRAD_wear-out}
During \mechanism{} operation, local PRAD is reprogrammed once \onurtwo{every} $1/\sigma_1$ writes on average (see Section~\ref{sec:random_remap_period}). Hence, local PRADs wear out at a rate $1/\sigma_1$  times slower than the memory array \onureight{(e.g., for $\sigma_1=1$\%, 100x slower)}. 
The global PRAD is reprogrammed every 1/$\sigma_2$ ($512\times100$) writes on average, which makes its wear out negligible compared to the wear out of the resistive memory cells. 

\section{Related Work}
\label{sec:related_work}


\onur{To our knowledge,} \mechanism{} is the first work that seamlessly integrates wear-leveling and fault tolerance techniques in the same mechanism. We have already discussed and evaluated Security Refresh~\cite{seong2010security} in Sections~\ref{sec:experiemntal_setup} and \ref{sec:evaluation}.
\onur{We now} briefly discuss other resistive memory techniques for \onur{enhancing} lifetime, \onur{wear-leveling} and fault tolerance.

\subsection{Wear-Leveling Techniques}
\vspace{5pt}\noindent\textbf{Wear-Leveling Techniques for PCM.}
There are many prior works that propose wear-leveling techniques to enhance PCM \onurfour{lifetime~\cite{zhou2009durable,YongsooJoo2010,Seznec2010,qureshi2009enhancing,seong2010security,qureshi2011practical,Yu2014,JoosungYun2012,Fan2014,han2015enhanced,zhou2016increasing,Zhang2017,Chen2012,Chang2016,Long2013,DuoLiu2014,Pan2016,AghaeiKhouzani2014,Im2014,Cheng2016}}. These works propose different techniques \onurtwo{to optimize} wear-leveling via swapping and remapping data.
Several prior works propose wear-leveling mechanisms that are aware of process variation across the memory chip~\cite{han2015enhanced,zhou2016increasing,Zhang2017}.
\onur{Several techniques use} OS support to improve PCM wear-leveling~\cite{Chen2012,Chang2016,Long2013,DuoLiu2014,Pan2016,AghaeiKhouzani2014,Im2014,Cheng2016}.

Unlike \mechanism{}, none of these works 
implement or discuss
how to integrate a fault \onur{tolerance} mechanism that works with the proposed wear-leveling techniques. Also, some of these techniques require storing and maintaining large remapping tables~\cite{Seznec2010,zhou2009durable}
, which can incur significant storage and latency overhead. 
  

\vspace{5pt}\noindent\textbf{Wear-Leveling Techniques for Hybrid DRAM/PCM Memory.} \lois{DRAM/PCM} hybrid memories \lois{aim} to \onur{provide} the best of both worlds: the low access latency of DRAM, and the large storage capacity of PCM. Existing wear-leveling techniques  1) minimize the number of writes by reducing the number of dirty evictions to PCM and re-compute results instead of saving data in PCM~\cite{Hu2011}, 2) use techniques to allocate \onur{heavily-written} data in DRAM only~\cite{Li2012,yoon2012row}, or 3) \onur{migrate} \onur{heavily-written} pages from PCM to DRAM~\cite{Zhang2009}. \mechanism{} can be combined with these techniques to further improve wear-leveling \onur{effectiveness}. 

\subsection{Fault Tolerance and \onur{Error} Correction}

\onur{There are many fault tolerance and error correction techniques that can be applied to resistive \onureight{memories~\cite{schechter2010use,qureshi2011pay,tavana2017remap,seong2010safer,fan2013aegis,melhem2012rdis,ipek2010dynamically,azevedo2013zombie,tavana2018block,yoon2011free,awasthi2012efficient,udipi2012lot,Li2011,ShengLi2012,li2019selective,qin2005safemem,capps1999method,patel2019understanding,patel2020bit,kline2020,nair2013archshield,kline2017sustainable,cai2015data,cai2015read,luo2018heatwatch,cai2014neighbor,cai2012flash,luo2018improving,cai2017vulnerabilities,cai2013program,cai2013threshold,cai2012error,luo2015warm,luo2016enabling,cai2018errors,cai2017error,yoon2010virtualized,yoon2010virtualized2,wilkerson2008trading,wilkerson2010reducing,alameldeen2011energy,khan2014efficacy}}.
Among these works, there are several that focus specifically on resistive \onurfour{memories~\cite{schechter2010use,qureshi2011pay,tavana2017remap,seong2010safer,fan2013aegis,melhem2012rdis,ipek2010dynamically,azevedo2013zombie,tavana2018block,yoon2011free}} that can be classified into four categories}. First, techniques that replace faulty cells with redundant cells~\cite{schechter2010use,qureshi2011pay,tavana2017remap}. 
Second, techniques that use data partitioning and \onurfive{inversion~\cite{seong2010safer,fan2013aegis,melhem2012rdis,Zhang2017-2}}. SAFER~\cite{seong2010safer}, Aegis~\cite{fan2013aegis}, RDIS~\cite{melhem2012rdis}, \onur{and Zhang et al.~\cite{Zhang2017-2}} exploit the observation that a stuck-at-value memory cell remains readable, 
and employ data partitioning and inversion of faulty partitions 
to tolerate cell failures. 
Third, techniques that use faulty page and block pairing. DRM~\cite{ipek2010dynamically} tolerates block failures within a page by pairing it with another page such that failed blocks do not intersect. Zombie memory~\cite{azevedo2013zombie} corrects errors in memory blocks by pairing them with working blocks of decommissioned pages. Block Cooperation~\cite{tavana2018block} repurposes faulty blocks to provide support to working blocks within the same page to keep \onur{the page} “alive”.
\onur{Fourth}, techniques that use ECC. \mbox{FREE-p}~\cite{yoon2011free} performs fine-grained remapping of memory blocks by storing  remapping pointers in the functional cells of a worn-out block. FREE-p protects against both hard and soft errors.
Unlike \mechanism{}, none of these mechanisms \onurtwo{consider} the integration of a \joaotwo{wear-leveling} \onurtwo{mechanism} with a fault tolerance or error correction mechanism, which is essential to make these mechanisms work in real systems.

\subsection{Other Lifetime Enhancement Mechanisms}
\noindent\textbf{Device-Level Techniques.} \onur{Several works use device-level techniques to improve resistive memory lifetime~\cite{Jiang2012,shevgoor2015improving}.} Jiang et al.~\cite{Jiang2012} propose using the 2-bit MLC cell as a tristate cell to reduce the RESET current to increase PCM endurance. This technique can be used together with \mechanism{} to further improve memory lifetime.

\vspace{5pt}\noindent\textbf{Reducing Redundant Writes.} Many prior works improve PCM lifetime by reducing the number of written bits into memory~\cite{Lee2009,zhou2009durable,YongsooJoo2010,Cho2009,Sun2011,lee2010phase}. Some works~\cite{Lee2009,zhou2009durable,YongsooJoo2010,Cho2009} propose writing to the memory array only those bits whose values had been changed in the row buffer, which reduces the number of cells modified on each write. 
\onureight{All these techniques can be used together with \mechanism{} to further improve memory lifetime.}


\section{Conclusion}

We propose \mechanism{}, the first mechanism that combines wear-leveling and fault tolerance seamlessly and efficiently \lois{in resistive memories}. \mechanism{} wear-leveling mechanism remaps writes on-the-fly to random locations, 
and \mechanism{} fault tolerance mechanism remaps \onurtwo{a failed memory block to an empty (unoccupied or spare) memory block}.

Unlike previous works, \mechanism{} integrates both \onur{mechanisms} efficiently using \joaotwo{a} \onur{programmable resistive} address decoder (PRAD), which decouples memory addresses from physical memory locations.
\mechanism{} \onur{enables} rapid leveling of the write non-uniformity, and fine-grained remapping of faulty memory blocks.

Our \lois{\onur{evaluations show} that \mechanism{}
\onur{combined with $\text{ECP}_1$ (error correcting pointers)}
\onur{provides} 68\% longer lifetime, and \mechanism{} \onur{combined with $\text{ECP}_6$ provides} 87\% longer lifetime, compared to the best state-of-the-art \joaotwo{wear-leveling} mechanism and fault correction techniques, for \onur{similar or less} area overhead. The average performance (energy) \onur{penalty} of \mechanism{} is 0.51\% (0.47\%), compared to a baseline system without any wear-leveling or fault \onur{tolerance} techniques.}
\onur{We conclude that \mechanism{} is an effective and low-cost reliability solution for resistive memories.}
\section*{Acknowledgements}

We thank \onurfour{the SAFARI Research \onurfive{Group} members for the valuable input and  \onurfive{the} stimulating intellectual environment they provide, the anonymous reviewers for their feedback, and the industrial partners of SAFARI, especially ASML, Google, Huawei, Intel, Microsoft, and VMware, \onurfive{for their support.}}


{
  \footnotesize
  \renewcommand{\baselinestretch}{0.5}
  \let\OLDthebibliography\thebibliography
  \renewcommand\thebibliography[1]{
    \OLDthebibliography{#1}
    \setlength{\parskip}{0pt}
    \setlength{\itemsep}{0pt}
  }
  \bibliographystyle{IEEEtranS}
  \bibliography{refs}
}

\end{document}